\documentclass[fleqn,12pt]{wlscirep}
\usepackage{color}
\usepackage{epstopdf}
\usepackage{todonotes}
\usepackage{setspace}
\usepackage{upgreek}
\usepackage{tabu}
\usepackage{booktabs}
\usepackage{siunitx}
\usepackage{lineno}
\usepackage{amsmath,amssymb}
\usepackage{calrsfs}
\DeclareMathAlphabet{\pazocal}{OMS}{zplm}{m}{n}
\usepackage{setspace}

\title{Ice-rule made manifold: phase transitions, topological defects and manifold restoration in two-dimensional artificial spin systems}

\author[1,*]{Gavin M. Macauley}
\author[1]{Gary W. Paterson}
\author[1,2]{Yue Li}
\author[1,3]{Rair Mac\^edo}
\author[1,*]{Stephen McVitie}
\author[4,*]{Robert L. Stamps}
\affil[1]{SUPA, School of Physics and Astronomy, University of Glasgow, Glasgow, G12 8QQ, United Kingdom}
\affil[2]{Materials Science Division, Argonne National Laboratory, Lemont, IL, 60439, USA}
\affil[3]{Electronics and Nanoscale Engineering Division, James Watt School of Engineering, University of Glasgow, Glasgow, G12 8QQ, United Kingdom}
\affil[4]{Department of Physics and Astronomy, University of Manitoba, Winnipeg, Manitoba, MB R3T 2N2, Canada}
\affil[*]{gavin.macauley@glasgow.ac.uk, stephen.mcvitie@glasgow.ac.uk, robert.stamps@umanitoba.ca}

\begin{abstract}

\textbf{Artificial spin ices are arrays of correlated nano-scale magnetic islands that prove an excellent playground in which to study the role of topology in critical phenomena. 
Here, we investigate a continuum of spin ice geometries, parameterised by rotation of the islands.
In doing so, we morph from the classic square ice to the recently studied pinwheel geometry, with the rotation angle acting as a proxy for controlling inter-island interactions.
We experimentally observe a change in ground state magnetic order from antiferromagnetic to ferromagnetic across this class of geometries using Lorentz transmission electron microscopy on thermally annealed cobalt arrays.
The change in ordering leads to an apparent change in the nature of the defects supported: from one-dimensional strings in the antiferromagnetic phase to two-dimensional vortex-like structures in the ferromagnetic one, consistent with the scaling predicted by the Kibble-Zurek mechanism. 
Our results show how magnetic order in artificial spin ices can be tuned by changes in geometry so that a truly frustrated ice-rule phase is possible in two-dimensional systems. Furthermore, we demonstrate this system as a testbed to investigate out-of-equilibrium dynamics across phases.}

\end{abstract}
\begin{document}

\flushbottom
\maketitle

\thispagestyle{empty}
\section*{Introduction}

Artificial spin ices (ASIs) are arrays of strongly correlated sub-micron magnetic islands in which the individual elements are coupled through magnetostatic interactions\cite{2006Wang}.
The aspect ratio of these elements is usually chosen so that they behave as single domain Ising macrospins. 
As the long-range dipolar field mixes spin and spatial degrees of freedom, their collective nature depends strongly on the exact arrangement and orientation of the nano-magnets \cite{1991Malozovsky,2006Politi}. 
Interactions can then be tuned locally, achieving experimentally tractable mesoscopic analogues to atomistic systems.

Ising-like systems made of magnetic islands have been studied extensively for several decades\cite{2000Cowburn}, and continue to merit attention for the unusual ordering created through tailoring the inter-element interactions\cite{2016Arnalds,2017Nguyen,2016Nisoli}. 
More recently, systems obeying Potts-like\cite{2019Sklenar, 2018Louis} and dipolar XY\cite{2018Streubel, 2018Schildknecht, 2018Leo} Hamiltonians have been manufactured from assemblies of patterned magnetic structures using precision lithography. 
The individual microstate of these systems can be interrogated using magnetic microscopy (e.g. MFM~\cite{2006Tanaka}, PEEM~\cite{2013Farhan} or Lorentz TEM~\cite{2011Daunheimer,2008Qi,2018Li}) in response to external stimuli such as applied field~\cite{2015Burn, 2018Li}, electrical current~\cite{2015Le,2017Jungfleisch} or temperature\cite{2017Gliga}.
ASIs thus prove new platforms in which to examine aspects of physics otherwise not directly observable. 
These include glassiness~\cite{2016Morley}; charge fragmentation~\cite{2016Canals}; and topologically-induced textures such as magnetic `monopoles'~\cite{2010Ladak, 2011Mengotti,2019Farhan} and Dirac strings~\cite{2016Vedmedenko}.

A new geometry which provides one such platform to explore these ideas is the pinwheel form of ASI. 
This structure is formed by rotating each island in the classic square ASI through $45^{\circ}$ about its centre, and has evoked recent interest for its dynamic chirality\cite{2017Gliga}; homogeneous domain wall reversal processes\cite{2018Li}; controllable anisotropy\cite{2018MacedoPRB}; and as a vortex channel in superconducting-ASI heterostructures\cite{2018Wang}.

In this work, we investigate the class of ASI geometries which covers the continuum between square and pinwheel ice.
Across these structures, a transition in spin ordering is predicted to emerge\cite{2018MacedoPRB}, from antiferromagnetism (AFM) in the classic square lattice\cite{2006Wang} to ferromagnetism (FM) in the pinwheel lattice.
As the intrinsic coercive field of an island is assumed the same in every geometry, this transition in ordering is due solely to the change in dipolar interactions.
We experimentally confirm the change in ground state and observe directly how this transition is mediated by different defect textures in the two phases using Lorentz transmission electron microscopy (LTEM) and in-situ annealing of cobalt (Co)  structures formed by focused electron beam induced deposition (FEBID).
We find certain tilings support mixed phases rather than an abrupt change in ordering.
This follows from competition between the dipolar interactions and the coercive field barriers, and leads to a quenching similar to that in idealised two-dimensional spin ice models~\cite{2012Levis}.
The cooling process then unavoidably establishes defects as the system orders, consistent with the Kibble-Zurek mechanism\cite{1985Zurek, 1976Kibble, 2013delCampo} (KZM).
This mechanism predicts how defects scale as a system is driven through a second order phase transition.
Notably, the change in tiling appears to affect the dimensionality of the defects.
In the AFM phase, defects take the form of one-dimensional (1-D) strings~\cite{2013Gliga, 2016Vedmedenko}, whereas in the FM phase, two-dimensional (2-D) structures appear, including many similar to vortices.
As a measure of the defect density (a common figure of merit for the KZM\cite{2013Ulm}), we characterise these by their circulation, and show numerically that they scale with cooling rate in accordance with the exponent predicted by the KZM for the 2-D Ising universality class.
As a corollary, we show that this class of geometries includes a frustrated ice-rule manifold in 2-D.

\section*{A class of geometries}
\begin{figure}[]
\centering
\includegraphics[width=0.875\textwidth]{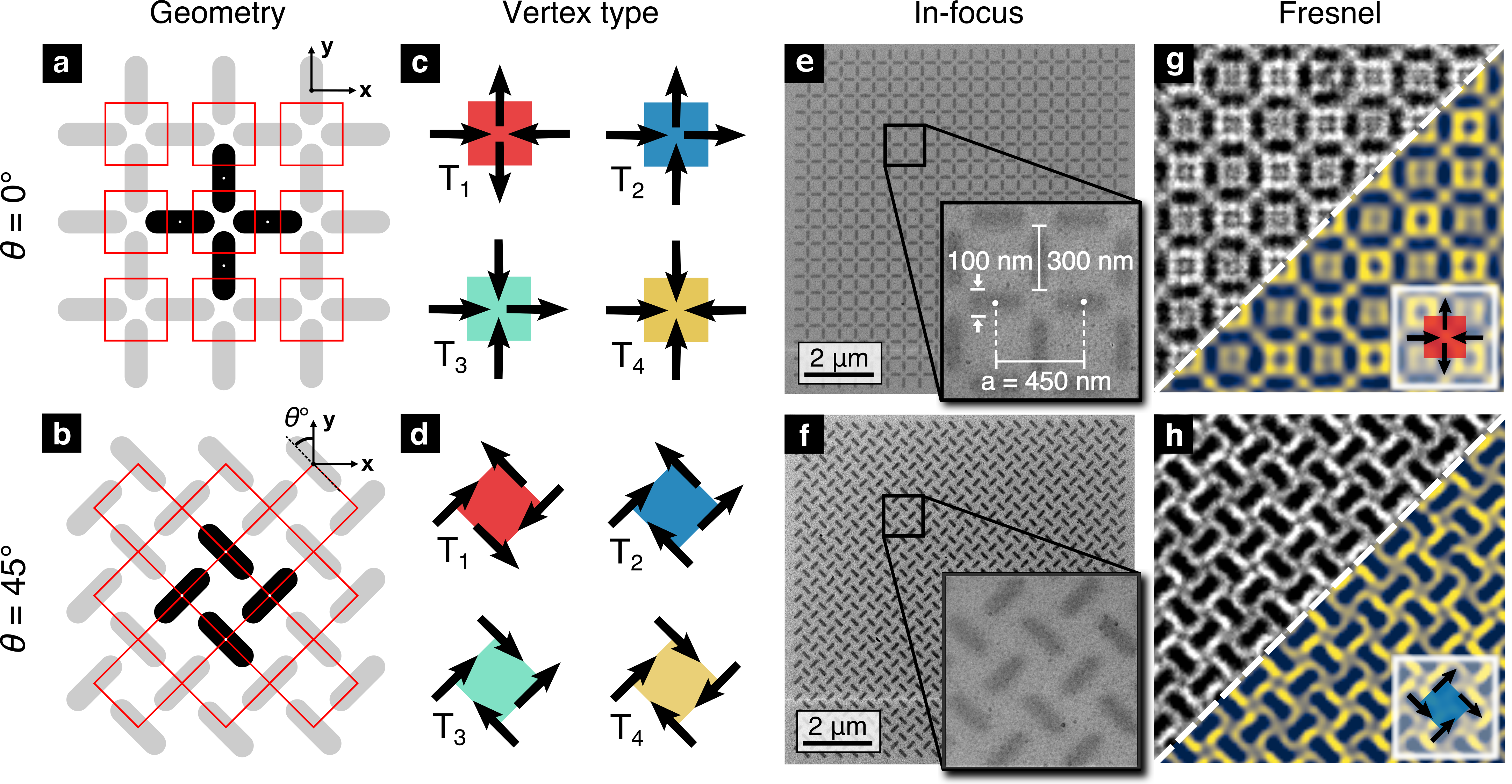}
\caption{\textbf{ A continuum of geometries defined by rotation angle.}
\textbf{a},\textbf{b}, 
The square ice tiling in \textbf{a} is transformed by rotating each island through $\vartheta$. 
A schematic of the  $\vartheta = 45^{\circ}$ case, pinwheel ice, is shown in \textbf{b}. 
For both, the location of vertices are highlighted in red. 
\textbf{c},\textbf{d},
An example of each of the four vertex types for \textbf{c}, $\vartheta = 0^{\circ}$, and \textbf{d}, $\vartheta = 45^{\circ}$  ice. 
A consistent colour coding for vertex type is used throughout this work.
\textbf{e},\textbf{f},
In-focus TEM images of FEBID arrays for these two angles; dimensions indicated in the inset to \textbf{e} are the same as those in \textbf{f} and for all fabricated arrays. 
\textbf{g},\textbf{h},
Fresnel images (raw in greyscale, Fourier-filtered in colour) obtained after annealing for square and pinwheel ice. The difference in contrast across the long axis of the island (dark edge compared with light edge) is used to identify the spin direction and, hence, vertex types.}
\label{geometry_results}
\end{figure}

Our base structure, the square ice tiling (Fig.~\ref{geometry_results}a), is formed by two interleaved, orthogonal sub-lattices of elongated uniformly aligned nanomagnets.
From this tiling, we can obtain a continuum of geometries through rotating each island by an angle, $\vartheta$, about its midpoint.
This rotation acts as a proxy for controlling interactions between classes of neighbouring spins, weakening the nearest-neighbour couplings which dominate the square lattice\cite{2018MacedoPRB}.
Consistent with previous work, we choose the zero of this rotation, $\vartheta = 0^{\circ}$, to be square ice, and term the $\vartheta = 45^{\circ}$ state `pinwheel ice' (Fig.~\ref{geometry_results}b). 

Square artificial spin ice is conventionally studied in terms of `vertices' where four islands meet head on (highlighted in red in Fig.~\ref{geometry_results}a,b).
The system can then be described in terms of a `gas' of interacting vertex moments, rather than through the full ensemble of island spins\cite{2010Nisoli}.
Given that the macrospin associated with each island can point in one of two directions parallel to its long axis, there are 16 possible vertices.
These are sorted into four distinct types, T$_{1}$-T$_{4}$, of increasing magnetostatic energy; an example vertex for each type in square and pinwheel\cite{note1} ice is given in Fig.~\ref{geometry_results}c and d respectively. 
We note that T$_1$ and T$_4$ vertices carry no net moment. 
In this regard, they are antiferromagnetic vertices, as the island moments align antiparallel within each sub-lattice.
In this nomenclature, the ground state configuration for square spin ice is AFM, as it is composed of a chequerboard pattern of the two possible T$_1$ vertices (discussed first in Ref. [\citen{2006Wang}], experimentally confirmed in Ref. [\citen{2010Morgan}]).
Since the T$_2$ and T$_3$ vertices possess a net moment, we consider them ferromagnetic vertices\cite{note2}. 

Rotating each island in a vertex changes neither its type nor its ordering so that, say, a T$_2$ vertex carries a net moment irrespective of angle. 
However, it does markedly affect the energies of the types.
In square ice, the four types possess four well-separated energy levels, whereas in pinwheel ice, they are nearly degenerate\cite{2018MacedoPRB}.
For any given $\vartheta$, the ordering in an array may be characterised in terms of its fractional population of vertex types, $n_i$.

In this study, we subject spin ice arrays of various rotation angles to a thermal annealing protocol and image the resulting configuration using Lorentz TEM to characterise the ordering.
All arrays possess four-fold symmetry, so we restrict $\vartheta$ to the interval $[0^{\circ}, 90^{\circ}]$.
Samples were fabricated using FEBID of the metal-organic precursor Co$_2$$($CO$)_8$. 
Insofar as possible, all Co arrays were deposited under the same deposition conditions (as described in Methods).
This ensures that the intrinsic energy barrier of each island is nominally the same across every tiling.
Differences in behaviour then stem solely from the nature of the inter-island interactions as controlled by $\vartheta$.
Each array covered $(10.8$ ${\upmu}$m)$^2$, comprising 840 islands (a lattice of at least $20 \times 20$ vertices). 
The lateral dimensions of the islands were $300$ nm by $100$ nm, with a lattice constant, $a = 450$ nm, as shown in Fig.\ref{geometry_results}e. 
The thickness of the islands over the central portion of each array was measured to be $\sim3$ nm using atomic force microscopy (see supplementary, \S1). 
Islands closer to the outer edges were thicker ($\sim4$ nm), consistent with the effects of gas diffusion during deposition\cite{2005Ding}. 
Similarly, magnetic contrast was poorer there, in line with there being lower local Co content and a greater deposit of carbon.
As such, statistics were drawn only from those macrospins which could be easily identified (in effect, the central portion of each array).
By considering only islands far from the array edges, we omit those topological defects which tend to form at the boundaries where the local field environment is different, and can disregard thicker islands which possess higher blocking temperatures.
This mitigates the effect of finite size and enables us to probe something approaching the bulk behaviour of each tiling pattern. 

Several arrays were deposited for each angle (at least four repetitions; six for those arrays in the range $[35^{\circ}, 55^{\circ}]$; with an additional two for the $\vartheta = 45^{\circ}$ case). The arrays were thin enough to be thermally active close to room temperature while still providing sufficient magnetic contrast for imaging in LTEM. 
In the Fresnel mode of Lorentz microscopy, contrast arises from deflection of the electron beam by the integrated magnetic induction of each island\cite{2006McVitie}. 
By comparing the intensity on either side of the long axis of each island in a defocused Fresnel image (Fig.~\ref{geometry_results}g,h), the macrospin orientation can be assigned.
The thermal annealing protocol involved heating the arrays to 250$^{\circ}$C \emph{in-situ} in a field-free environment ($\leq 0.1$~Oe). 
This was above the blocking temperature of individual Co islands, such that their associated Ising spin was superparamagnetic.
This temperature was maintained for two hours before the arrays were cooled at a rate of $1.5^{\circ}$C min$^{-1}$ to $\sim -10^{\circ}$C. 

For the range of samples studied, the onset of island flipping occurred at $110^{\circ}$C, and continued until 250$^{\circ}$C. 
This range reflects the fact that blocking temperatures are site dependent. 
We assume a saturation magnetisation, $M_S$, equal to  70$\%$ that of bulk Co (consistent with the Co content obtained under similar deposition conditions\cite{2010Cordoba, 2011Belova}).
Together with the island dimensions above, the characteristic energy scale in terms of the dipolar constant, $D = \mu_0 (M_S V)^2/(4\pi a^3) = 0.033$ eV, where $V$ is the volume of the island.  
This corresponds to a temperature of $118^{\circ}$C---extremely close to the observed onset of flipping. 
After cooling, Fresnel images were taken and analysed with the aid of semi-automatic image processing to extract the direction of magnetisation of each island. This is further discussed in supplementary, \S2.

 \begin{figure}[]
\centering
\includegraphics[width=0.95\textwidth]{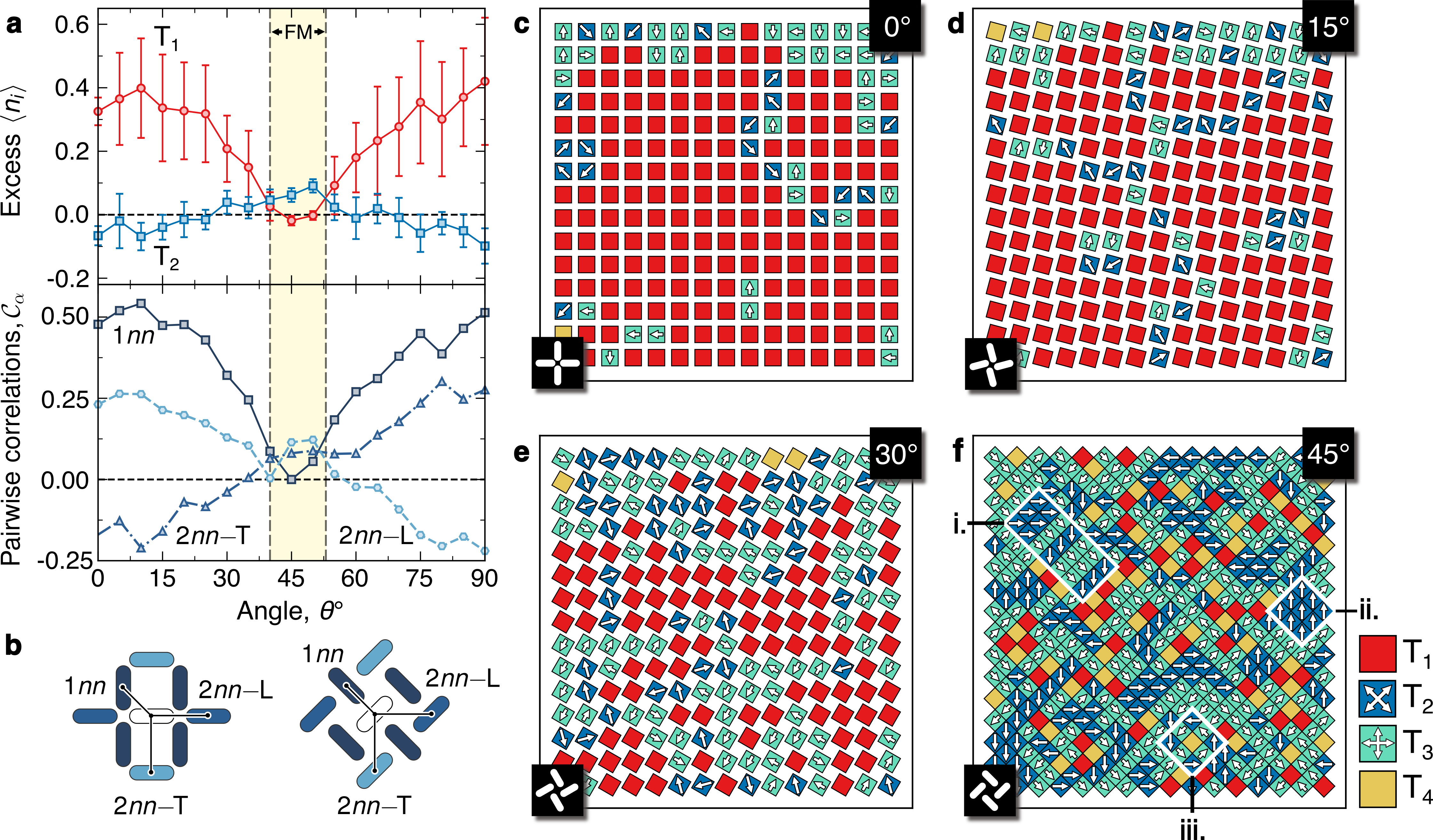}
\caption{\textbf{A transition in ground state ordering with island rotation.} 
\textbf{a},\textbf{b}, The upper panel of \textbf{a} shows the change in excess fractional populations of T$_1$ and T$_2$ vertices with angle. 
Error bars reflect $\pm$1~standard deviation when the data is averaged over all samples. 
The lower panel of \textbf{a} displays the correlations for three classes of near neighbours labelled by the schematic in \textbf{b}. For both graphs, the dashed horizontal line refers to the expected statistics for an uncorrelated sample. 
The shaded region highlights the FM phase i.e. the angular region for which T$_2$ vertices are in excess above T$_1$ vertices.
Full populations and uncertainties in the correlations are given in supplementary, \S3, Figs.~S5 and S6. 
\textbf{c}-\textbf{f}, Typical annealed configurations in terms of vertex type for $0^{\circ}$, $15^{\circ}$, $30^{\circ}$, and $45^{\circ}$ tilings, respectively. 
Each 15$\times$15 configuration is drawn from the central `best contrast' portion of an experimental array. 
Each square represents a four-island-unit in the array as depicted in bottom left hand corner of each panel. 
The vertex moment (magnitude and direction) is shown using arrows for FM vertices.
A full rotation series is given in supplementary, \S3, Fig.~S7.}
\label{ordering_results}
\end{figure}

In Fig.~\ref{ordering_results}, we present the results of the thermal annealing. 
The upper panel of Fig.~\ref{ordering_results}a shows the change in excess\cite{2006Wang} populations of T$_1$ and T$_2$ vertices with angle, with reference to a perfectly uncorrelated sample.
We pick T$_1$ and T$_2$ vertices as they function as an `indicator species' for the AFM/FM ordering within an array. 
Full populations are given in supplementary, \S3, Fig.~S5, but we note that the remaining high-energy types, T$_3$ and T$_{4}$, are always suppressed with respect to the expected statistics for an uncorrelated sample. 

Consistent with the experimentally established ground state of square ice, there is a strong excess of T$_1$ vertices near to $0^{\circ}$.
Comparing tiling patterns from $0^{\circ}$ to $90^{\circ}$ in steps of $5^{\circ}$, this population decreases to a minimum at $45^{\circ}$ before becoming maximal once more at $90^{\circ}$, as expected for square ice. 
Importantly, there exists a region from approximately $39^{\circ}$ to $53^{\circ}$ where the ferromagnetic T$_2$ vertices are in \emph{slight} excess, consistent with previous predictions\cite{2018MacedoPRB}. 
We make clear that this transition to a T$_2$-excess regime is an approximation to the phase transition expected in the thermodynamic limit, as the experimental observation relies on finite-sized arrays.
Indeed, later in the paper, we make use of the fact that the transition persists in Monte Carlo (MC) simulations of pseudo-infinite arrays.
For simplicity, we refer to both scenarios---finite-sized arrays and the thermodynamic limit---as exhibiting a transition in ordering with angle.

Two further features are worth remarking upon. 
First, the transition from the T$_1$-excess phase to the T$_2$-excess phase is gradual, rather than abrupt. 
This suggests the possibility of observing mixed AFM/FM ordering. 
At the transition angles where the populations satisfy $n_1 = n_2$, the ice manifold is recovered i.e. all two-in-two-out vertices are equally likely. 
Achieving this equivalence has previously relied on introducing a height offset between sub-lattices\cite{2006Moller, 2016Perrin,2019Farhan} or by coupling XY mesospins to the square tiling\cite{2018Ostmanintmod}.
Here, we have restored the degeneracy of near-neighbour interactions through modifying the orientations of the nano-magnets.
The transition angles then represent planar geometries which obey the ice-rule fully.
Secondly, the signal from the FM phase is generally weak: the excess T$_2$ population is $\sim10\%$ in the FM phase when averaged over all samples. 
This will become an important theme in our discussion of defect formation in the FM phase later in the paper.

The lower panel of Fig.~\ref{ordering_results}a encapsulates these observations in terms of pairwise correlations between island spins. 
We use the correlation functions as defined by Ref.~[\citen{2006Wang}]: namely, a pair of islands contribute~$+1$~($-1$)~if the two moments are aligned in such a way as to minimise (maximise) the corresponding point dipolar energy. 
If the dipolar energy of a pair is zero, it contributes $0$ to the correlation function. 
We distinguish among three classes of neighbours depicted in Fig.~\ref{ordering_results}b: first-nearest-neighbours (1\emph{nn}); second-nearest-neighbours which are separated \emph{laterally} i.e. in the same vertex (2\emph{nn}-L); and second-nearest neighbours which are separated in a \emph{transverse} fashion i.e. in adjacent vertices (2\emph{nn}-T).  In pinwheel ice, these two classes of second nearest-neighbours are equivalent.

In general, the 1\emph{nn} correlations are strong in the AFM phase and suppressed in the FM phase. 
These strongly-coupled 1\emph{nn} pairs drive the AFM ordered ground state of square ice, whereas the FM ordering is stabilised by the relative strength of the 2\emph{nn} correlations compared with the 1\emph{nn} correlation for $\vartheta  \in \big[ 39^{\circ}, 53^{\circ} \big]$. 
At the transition angles, the arrays approximate well a truly frustrated system and the correlation functions are almost all equal, consistent with the system supporting both AFM and FM ordering. 

Figs.~\ref{ordering_results}c-f present typical vertex configurations obtained after annealing $0^{\circ}$, $15^{\circ}$, $30^{\circ}$, and $45^{\circ}$ arrays, respectively. 
Here, each four-island vertex is represented by a colour-coded square.
An arrow is superimposed on those vertices which possess a net moment.
Each panel displays $15 \times15$ vertices, reflecting the fact that the data has been `cut' to remove spins near to the edges of the arrays.
For the square ice case (Fig.~\ref{ordering_results}c), we observe the formation of large domains of T$_1$ separated by `string' defects i.e. a state where the ordering appears relatively long-ranged in terms of ground state clusters. 
For the pinwheel ice case (Fig.~\ref{ordering_results}f), the majority of vertices are FM and a variety of structures is seen, including stripe patterns, small domains, and flux closure regions (marked i., ii., and iii. respectively).
However, the ordering is short-ranged, extending only a few lattice constants.
Comparing arrays with different rotation angles (in effect, moving from Fig.~\ref{ordering_results}c-f), we see a shrinkage of the T$_1$ area and an increase in FM coverage.
This is not an abrupt transition; instead, the long-range AFM phase breaks down gradually to be replaced by a short-range FM phase. 

\section*{Quench behaviour}

To understand the decrease in the correlation length with angle (quantified in supplementary, \S4), we consider the effect of the relative magnitude of inter-island interactions on the likelihood of single macrospin flips.
This shows that the rotation angle determines the timescale needed to establish equilibrium.
In general, the arrays approach their ground state by making a number of Ising spin flips, where a given flip may or may not be energetically favourable in global terms.
Each spin flip is, however, driven by some local fluctuation in energy, $\Delta E (\vartheta)$, on a scale set by the interactions in the system and, hence, indirectly by $\vartheta$.
This fluctuation allows the spin to overcome the intrinsic energy barrier to switching, $E_{\text{b}}$. 
For simplicity, we take $E_{\text{b}}\gg\Delta E (\vartheta)$, to be the same for every island and in every geometry.
Assuming a N\'eel-Arrhenius law for the switching behaviour\cite{2012Nisoli}, we can compare the spin flip rates, $\Gamma$, in the AFM phase (typified by the $0^{\circ}$ case) to that in the FM phase (typified by the $45^{\circ}$ case) through the ratio
\begin{linenomath*}
\begin{equation}
\frac{\Gamma_{0^{\circ}}}{\Gamma_{45^{\circ}}} = \frac{\tau e^{ -\beta ( E_{\text{b}} - \Delta E(0^{\circ}) )}}{\tau e^{ -\beta ( E_{\text{b}} - \Delta E(45^{\circ}) )}}  = e^{\beta (\Delta E(0^{\circ})- \Delta E(45^{\circ}))},
\label{neel_equation}
\end{equation}\end{linenomath*}
where $\tau$ is an attempt frequency, and $\beta = 1/(k_B T)$ is the inverse thermodynamic energy. 
In arriving at eq.~(\ref{neel_equation}), we have neglected any angular dependence in the prefactor.
In a fuller treatment, the attempt frequency would be affected by the diminution in interaction strengths with $\vartheta$.
For example, studies using harmonic transition state theory have shown that symmetry reduction can lead to a signifcant decrease in $\tau$ when comparing one and two ring kagome ASI arrays\cite{2017Liashko}.
In fact, when calculating both $\Delta E$ and $\tau$ in this scheme, the change in the prefactor can have a larger effect on the rate than the change in the activation energy.
This may be relevant for FM array geometries where different magnetic configurations are almost degenerate\cite{2018MacedoPRB}, and there exists the likelihood of a return to the initial state after subsequent spin flips.
Nonetheless, approximating the islands as point dipoles, the largest interaction strength in square ice  (originating from the first-nearest-neighbours, $J_{1\text{nn}} = 3\sqrt{2} \,D$) is six times greater than the largest interaction energy in pinwheel spin ice (originating from the third-nearest-neighbours, $J_{3\text{nn}} = 1/\sqrt{2} \,D$). 
Associating the size of the fluctuation with the largest interaction term, the ratio of the rates at $250^{\circ}$C should go as $\Gamma_{0^{\circ}}/\Gamma_{45^{\circ}} \approx 10$. 
This estimate suggests that pinwheel ice requires approximately one decade of time more to undergo the same number of spin flips as does square ice. 

\begin{figure}[]
\centering
\includegraphics[width=0.75\textwidth]{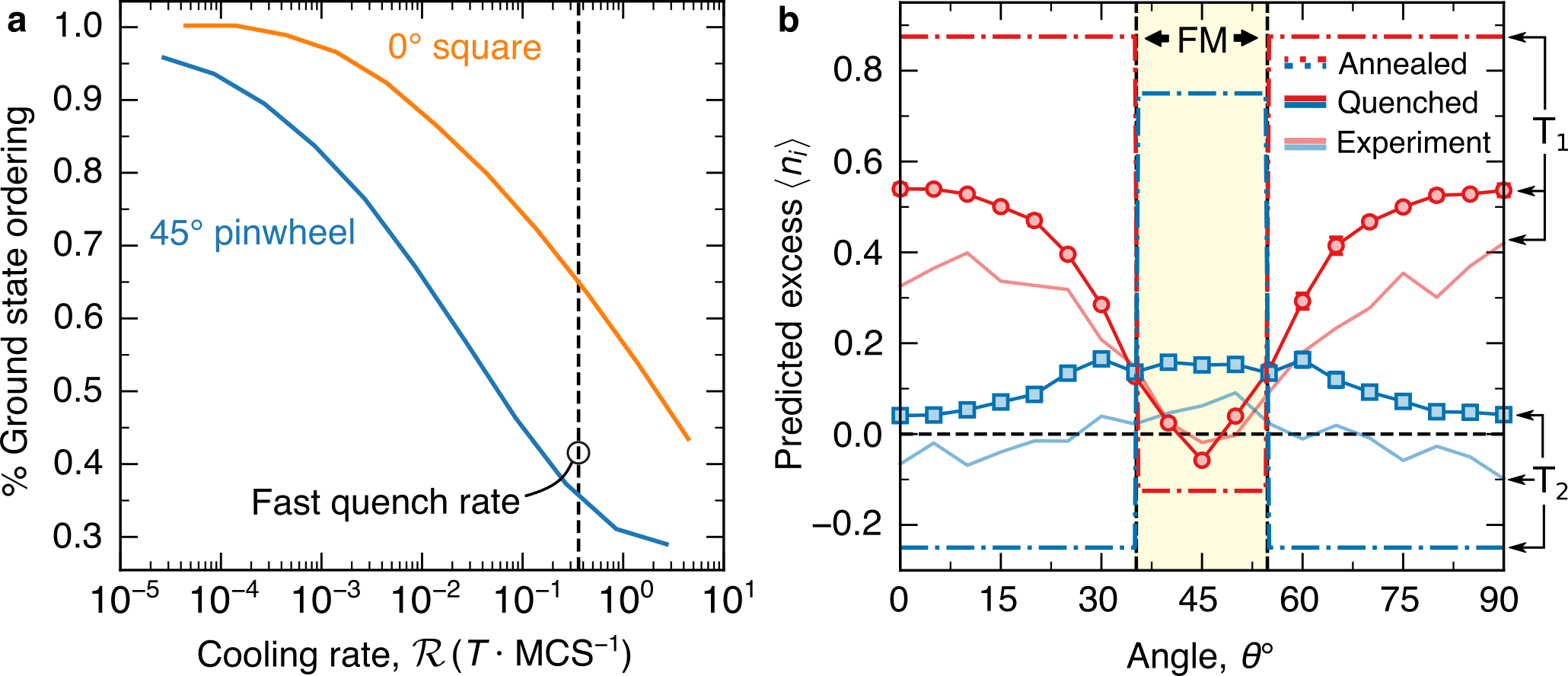}
\caption{\textbf{Statistics for quenched samples from Monte Carlo simulations.}
\textbf{a}, 
Relative ground state coverage as a function of the rate at which the samples are cooled from just above $T_c$ for infinite (i.e. with periodic boundary conditions) $0^{\circ}$ and $45^{\circ}$  tilings.
The dashed vertical line reflects a fast cooling rate which recovers well the experimentally obtained populations for the two tilings.
In general, the $45^{\circ}$ tiling lags the $0^{\circ}$ tiling in ground state coverage.
\textbf{b}, Expected vertex populations from MC simulations for both quenched (solid) and perfectly annealed (dashed-dotted) cases across the full angular range (all evaluated at the quench rate marked out \textbf{a}). 
The perfectly annealed samples show abrupt transitions between AFM and FM ordering. 
In the quenched case, these transitions are smeared out, consistent with the experimental populations of Fig.~\ref{ordering_results} (shown here by the faded lines). 
MC simulations performed for $50 \times 50$ vertex arrays with PBC.
The angular FM region depends on system size and so the highlighted region in \textbf{b} is broader that in Fig.~\ref{ordering_results}a, but consistent with that in Ref.~[\citen{2018MacedoPRB}].}
\label{quench_results}
\end{figure}

Fig.~\ref{quench_results} explores this in the context of Monte Carlo simulations.
In Fig.~\ref{quench_results}a, we show the percentage of ground state coverage for square and pinwheel ice as a function of the rate at which the systems are cooled from the high-temperature Ising paramagnetic phase.
The systems are initialised above their ordering temperatures, and then the temperature is decreased to zero in a variable number of steps; this acts as a proxy for cooling rate.
At each temperature point, one MC step is performed and the final vertex population at  $T = 0$ is recorded.
In the limit of an infinite number of steps, the arrays should explore all phase space\cite{note6} and find the true ground state, equivalent to cooling the systems adiabatically. 
We identify the change in temperature per MC step with a cooling rate, $\pazocal{R}$, an approach comparable to that in Ref.~[\citen{2014Liu}].
Pinwheel ice needs to equilibrate for longer to reach a similar coverage of ground state vertices and thus lags behind square ice by approximately one decade.

The experimentally weak excess in the numbers of T$_2$ vertices in the FM phase (Fig.~\ref{ordering_results}a) reflects the fact that the FM tilings are quenched to a greater extent than AFM tilings by the same anneal process.
Fig.~\ref{quench_results}b demonstrates this by plotting the excess populations as predicted by MC simulations of T$_1$ and T$_2$ vertices over the full range of angles for two cases: perfectly annealed infinite systems (dashed-dotted line) and a rapid quench (solid line with markers) corresponding to the rate marked in Fig.~\ref{quench_results}a.
The perfectly annealed simulations show an abrupt transition between AFM and FM ordering.
On the other hand, the quenched simulations---which are purposely not allowed to equilibrate at each temperature step---show excess populations in good agreement with those obtained in the experiment. 
In particular, there exists only a weak excess in the number of ground state vertices in the FM phase.

At this point, we emphasise the distinction between the two transitions that these geometries exhibit.
The first concerns the nature of the ground state and is controlled by $\vartheta$ i.e. it is a non-thermal transition driven by a geometrical parameter.
Secondly, each tiling undergoes a second-order phase transition as the blocking temperature is traversed.
In the experiment, a cooling rate of 1.5$^{\circ}$C~min$^{-1}$ resulted in a non-uniform degree of ordering for different island rotations.
This is a consequence of the fact that relaxation times differ vastly with rotation angle, and so different arrays were out of equilibrium to different extents.
This is a specific observation of a more general result: the dynamics of a system cease to be adiabatic in the vicinity of a critical point as the relaxation time diverges.
The non-adiabatic cooling enforces regions of incommensurate symmetries, separated by topological defects---compare, for example, He$^{4}$ quenched across the superfluid transition in which quantised vortex defects appear\cite{1994Hendry} or, analogously, the proposed role of cosmic strings in seeding galaxy formation in the early universe\cite{1993Zurek}.
The Kibble-Zurek mechanism describes the universal scaling laws underpinning the formation of such defects with cooling rate (discussed in Methods).
Our experiments allow us to investigate this mechanism in the context of ordering in artificial spin ices.

\section*{Defect formation}

\begin{figure}[]
\centering
\includegraphics[width=0.8\textwidth]{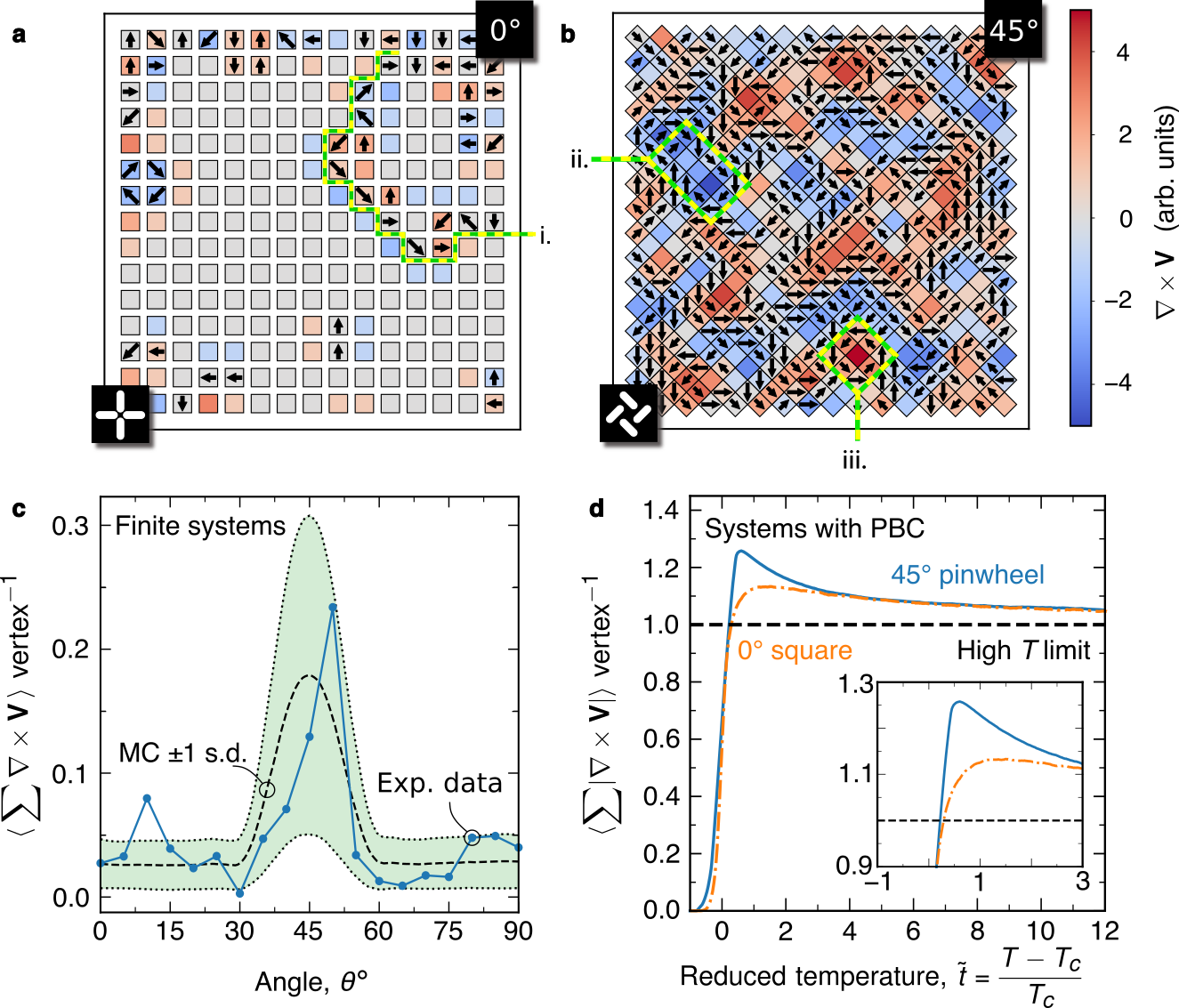}
\caption{\textbf{Vortex defects in the FM phase.} \textbf{a,b}, Curl maps corresponding to the experimentally obtained vertex configurations in Fig.~\ref{ordering_results}c,f (square and pinwheel ice). Topological defects (strings, i., in square ice; vortices, ii.-iii., in pinwheel ice) are highlighted. \textbf{c}, For finite systems, we integrate the curl over the area of the array as a measure of the net circulation. The shaded region highlights $\pm 1$~standard deviation around the mean MC statistics (dashed line) for this integrated curl. Both the experimental results (filled markers) and MC statistics for quenched finite-size arrays prove that these excitations impart a net circulation in the FM phase. \textbf{d}, MC simulations only: the integrated absolute value of the curl with temperature shows that this holds true even when the arrays are heated slowly. There exists a region around $\tilde{t} = 0$ for which both square and pinwheel ice rapidly produce curl, but this feature is more pronounced in the FM phase.}
\label{melting_results}
\end{figure}

It is well established that ordering in square ASI can occur by the formation of strings---either excited T$_2$ vertices on a ground state T$_1$ background\cite{2008Mol}, or low-energy T$_1$ vertices on a polarised T$_2$ lattice\cite{2012Kapaklis}.
Indeed, these topological defects are present in our annealed AFM configurations (Fig.~\ref{ordering_results}c). 
In the FM phase, by contrast, we observe a melting transition mediated by the formation of two-dimensional structures and, in particular, 2-D vortices.
These vortex structures are composed of nearby vertices, such that the vertex moments circulate around a central core. 
In analogy with the Bloch point in magnetic thin films, the role of this core is played by T$_{1}$/T$_{4}$ vertices, both of which have no moment.
For angles near to $45^{\circ}$, the moments in a T$_{4}$ vertex are arranged similarly to a vortex, while the moments in a T$_{1}$ vertex are akin to an anti-vortex (see supplementary, \S5, Fig. S9).

To quantify the emergence of a vortex regime in the FM phase, we treat the lattice of vertex moments as a discrete vector field, $\mathbf{V}$, and calculate its curl through a finite difference scheme to obtain a measure of the local circulation at each vertex\cite{note5}.
As the vertex moments are constrained to lie in-plane, only the z-component of $\nabla \times \mathbf{V}$ is non-zero and the curl may be regarded as a scalar field. 

In Fig.~\ref{melting_results}a,b, we display heat maps of $\nabla \times \mathbf{V}$ corresponding to the experimentally-obtained vertex configurations in Fig.~\ref{ordering_results}c,e  for $0^{\circ}$ square ice and $45^{\circ}$ pinwheel ice, respectively.
The curl is plotted on the same scale across both plots with the net moment shown atop FM vertices as before.
Comparing the two plots, the curl is significantly more pronounced in the $45^{\circ}$ case and traverses larger extremes.
Instances of topological defects are highlighted for both tilings; for example, i. in Fig.~\ref{melting_results}a labels a Dirac string within square ice.
In pinwheel ice, the defects are no longer one-dimensional chains but instead two-dimensional; ii. and iii. in Fig.~\ref{melting_results}b mark out instances of vortices with a clockwise and counter-clockwise circulation, respectively.
In both examples, FM T$_2$/T$_{3}$ vertices surround an inner core of T$_{1}$/T$_{4}$ vertices.
These cooperative structures constitute extrema of $\nabla \times \mathbf{V}$. 

Integrating $\nabla \times \mathbf{V}$ over the area of an array then gives a measure of the net circulation.
Typical results taken from our experimental data are shown in Fig.~\ref{melting_results}c. 
The solid line is taken from experimental data, and the dashed line refers to the mean statistics expected from MC simulations of quenched finite-size arrays.
The shaded region highlights $\pm 1$ standard deviation around this mean.
Excellent agreement between the experimental and Monte Carlo data is seen: both show a clear peak near to $45^{\circ}$ when the system begins to acquire a net circulation as a result of the preference for FM vertices and the tendency of the system to minimise stray field through forming short length scale vortices. 

This observation is consistent with the idea that the FM phase is dominated by vortex structures. 
On its own, however, it is not conclusive proof of a vortex-dominated melting/freezing \emph{transition}.
To show this, we consider MC simulations of the two prototypical angles, $0^{\circ}$ and $45^{\circ}$, as a function of temperature in Fig.~\ref{melting_results}d.
Here, we do \emph{not} quench the systems but quasi-adiabatically increase the temperature from $T=0$.
In both cases, the systems are initialised in their respective ground state.
The systems are equilibrated at each temperature point with the difference between consecutive temperatures of $10^{-5}\,D/k_b$ which is approximately $\sim$ 4 mK in units appropriate to our arrays.
For systems with periodic boundary conditions (PBCs), the integrated curl is no longer an appropriate figure of merit as neither tiling acquires a consistent net circulation in the thermodynamic limit. 
Instead, we chose the integrated absolute value of the curl.
In the paramagnetic limit, this quantity tends to $\approx 1$ per vertex\cite{note4}.
In the low temperature limit, the ground state of square ice (uniform T$_1$ tiling) and pinwheel ice (uniform T$_2$ tiling) have $\langle \sum | \nabla \times \mathbf{V} | \rangle = 0$. 
In Fig.~\ref{melting_results}d, we plot the evolution of this quantity with temperature. 
To make a comparison between square and pinwheel ice, we work in reduced units, $\tilde{t}  = (T-T_{C})/T_{C}$, in terms of the appropriate critical temperatures, $T_{C}(\vartheta)$, for each tiling.

Near to the ordering transition at $\tilde t = 0$, both systems undergo a period in which curl is rapidly generated. 
However, this feature is more pronounced and the integrated curl attains a greater magnitude in the FM phase. 
As the system melts to a disordered phase, 2-D structures---in particular, vortices---proliferate, spoiling the long-range ordering (supplementary, \S6,~Fig.~S11).
We emphasise that this vortex-mediated regime persists even when the temperature is changed slowly, as in Fig.~\ref{melting_results}d.
This suggests that these defects play a more general role in establishing magnetic order during melting and freezing.
Furthermore, their existence in MC simulations even when PBCs are applied implies that they can nucleate anywhere within arrays and do not simply migrate from edges.
As further evidence, supplementary, \S6,~Figs.~S10~and~S11 give spatially resolved vertex maps for square and pinwheel ice at various temperatures, as taken from MC simulations.

\begin{figure}[]
	\centering
	\includegraphics[width=0.45\textwidth]{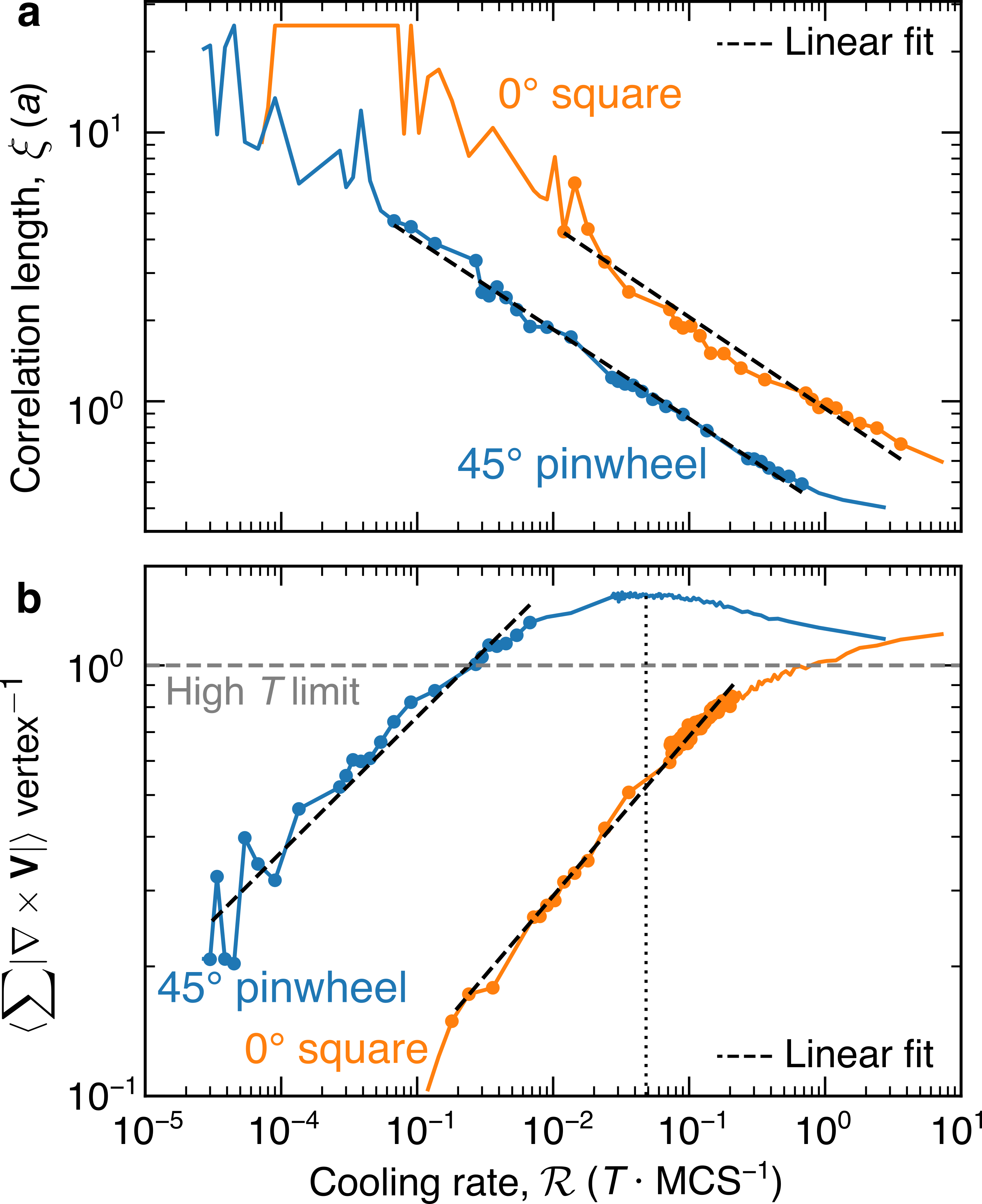}
	\caption{\textbf{Scaling of correlation length and defect density with cooling rate.}
	\textbf{a}, Correlation length, $\xi$, in units of the lattice constant, $a$, as a function of cooling rate for square and pinwheel ice. Taken from Monte Carlo simulations of $50 \times 50$ vertex arrays with PBC.
	\textbf{b}, As in \textbf{a} but for the appropriate defect density, $\langle \sum | \nabla \times \mathbf{V} | \rangle$. A maximum in the curl density is observed for pinwheel ice at the rate indicated by the dotted line. This corresponds to a freeze-out rate above which the dynamics of the system remain frozen. The linear portion of each series is used for fitting (points explicitly shown with markers; fits are shown in black dashed lines). Here, cooling rate is measured in units of temperature, $T$, per Monte Carlo step.}
	\label{KZM_results}
\end{figure}

To complete this discussion, Fig.~\ref{KZM_results} examines the correlation length, $\xi$, as extracted from the two point correlator, $\pazocal{G} (r) = \langle \vec{s}_i \cdot \vec{s}_j \rangle$, and the defect density, $\langle \sum | \nabla \times \mathbf{V} | \rangle$, as a function of cooling rate, $\pazocal{R}$, for square and pinwheel ice using MC simulations.
We have \emph{chosen} $\nabla \times \mathbf{V}$ as a measure of the defect density as it appears a natural choice for the vortices which form in FM arrays.
It is not necessarily as good a measure in the AFM phase where another quantity---string length, perhaps---would be more appropriate.
Strictly, the KZM is a statement of the expected power law dependence of the defect density with the `speed' at which the phase transition is traversed.
Attributing it to a system undergoing a phase transition relies on the values of the equilibrium critical exponents, and the scaling behaviour of the correlation length near $T_c$.
Coherent X-ray scattering provides quantitative evidence that square ASI belongs to the 2-D Ising universality class\cite{2019Sendetskyi} and, assuming this holds true for any rotation angle, the correlation length and curl should scale as $\pazocal{R}^{\pm 0.315}$ (see Methods for further justification). 
The integrated curl in pinwheel ice exhibits a peak (dotted vertical line, Fig.~\ref{KZM_results}b). 
We relate this feature to the freeze-out behaviour predicted by the KZM. 
Here, there appears a maximum cooling rate beyond which pinwheel ice cannot respond to changes in temperature.
For fast cooling rates, the curl stays close to the high temperature limit, and even vortices appear frozen out.

We perform a least squares fit to the linear portion of each series in Fig.~\ref{KZM_results} and extract the exponents (values $\pm1$ standard deviation are listed in Table \ref{table_kzm}).
All values are close to the predicted one for the 2-D Ising universality class.
We emphasise that other universality classes would predict different scaling exponents.
For example, it would be natural to draw an analogy between the formation of vortex structures in the FM arrays and those vortices which depin during a Kosterlitz-Thouless transition.
However, the exponential behaviour of the correlation length even at the critical point in a Kosterlitz-Thouless transition results in a more complex dependence on $\pazocal{R}$, i.e. one which is not simply some form of power-law.
Even assuming the asymptotic values of the critical exponents\cite{1998Luo} in this case (an approach which neglects key aspects of the KZM as applied to the Kosterlitz-Thouless transition\cite{2014Dziarmaga}) would give an estimate for the scaling exponent of~$\sim 0.5$.
Thus, our work suggests that both AFM and FM geometries obey the KZM consistent with the 2-D Ising universality class.

\begin{table}[]
	\centering 
	
	\begin{tabular}{l c c }
		\toprule & \multicolumn{2}{c}{Scaling Exponent}  \\ \cmidrule(lr){2-3} 
		& \centering{$\xi $}       & $\langle \sum | \nabla \times \mathbf{V} | \rangle$        \\ \toprule
		\midrule
		$\vartheta = 0^{\circ}$           &     $-0.340 \pm 0.005$        &  0.370 $\pm$ 0.004     \\
		$\vartheta = 45^{\circ}$        &       $-0.332 \pm 0.018$      &     0.314 $\pm$ 0.017   \\
		
		\bottomrule
	\end{tabular}
\caption{\textbf{Best fit critical exponents for $\xi$ and $\langle \sum | \nabla \times \mathbf{V} | \rangle$ associated with cooling rate, $\pazocal{R}$.} 
	Extracted from the linear portion of each series in Fig.~\ref{KZM_results} for $0^{\circ}$ square and $45^{\circ}$ pinwheel ice. 
	Assuming only the KZM and the equilibrium critical exponents of the 2-D Ising model, these quantities are predicted to scale with  $\pazocal{R}$ as  $\sim \,\pm 0.315$. 
	Uncertainties refer to $\pm 1$~standard~deviation in the least squares fit used.}
\label{table_kzm}
\end{table}

\section*{Discussion}

In this work, we have realised experimentally a system in which modifications in the lattice topology control the effective dipolar coupling between islands and act to tune the ground state between AFM and FM. 
Values of $\vartheta$ near to the transition give rise to competition between phases, while the transition angle itself corresponds to a two-dimensional manifold which satisfies the ice rule exactly.
Accessing this ice-rule phase in a completely planar system opens up the possibility of probing its frustrated dynamics and correlated interactions.
In addition, arrays comprised from a mixture of tiling patterns---square ice, say, joined to pinwheel ice---would offer the opportunity to study phase coexistence at interfaces and, even, engineer an ASI analogue of exchange bias.

Beyond the change in ground state, our thermal annealing reveals a transition in effective dimensionality of defects: from one-dimensional chains in the AFM phase to two-dimensional vortices in the FM phase.
Defect densities appear consistent with the Kibble-Zurek mechanism but a systematic  investigation of their experimental scaling with cooling rate would provide conclusive proof. 
These results illustrate the interplay between topology and magnetic order in artificial spin structures, and allow for the exploration of critical phenomena in frozen and glassy spin systems\cite{2016Morley}.
Our work demonstrates that this class of ASI is an exemplary testbed in which to probe both out-of-equilibrium dynamics and competing theories for melting across phases in low-dimensional many-body systems.\\

Original data files are available at DOI xx.xxxx.

\section*{Method}

\textbf{Sample Fabrication.} 
Arrays were written on a dual column Helios Plasma Focused Ion Beam instrument with gas injection system using electron beam induced deposition. 
Samples were deposited on $\sim$~40~nm thick electron transparent Si$_3$N$_4$ membranes for TEM measurement. 
Carbon was sputtered onto the membranes before Co deposition and a thin FEBID film applied after to inhibit oxidation and charging. 
The SEM beam current was $0.69$~nA; the accelerating voltage was $5$~kV; and the half-screen window size was $20.7$~$\upmu$m (magnification: 10000$\times$). 
The working distance was $4$~mm.
The gas flux of the precursor would tend to decrease over the course of a deposition session.
To compensate for this, arrays were deposited in a different order during subsequent sessions so that the average thickness remained constant across arrays.\\

\textbf{LTEM Measurement.} 
Annealing experiments were carried out \emph{in-situ} in a JEOL ARM200cF TEM equipped with a cold field emission gun operated at 200~kV. 
The beam spot size was 2; the emission current was 14~pA; and a 70~$\upmu$m condenser aperture was used. 
The objective lens was nulled before sample insertion to ensure a field-free environment ($\le$~1~Oe). 
A Gatan HC5300 holder was used to heat the arrays to 250$^{\circ}$C. 
This temperature was maintained for two hours to allow the arrays to equilibrate.
The samples were cooled to $\sim -10^{\circ}$C at 1.5$^{\circ}$C~min$^{-1}$, and Fresnel images taken using a Gatan Orius SC1000A CCD camera. 
These were processed to extract the macrospin orientation of each island. \\

\textbf{Monte Carlo.} 
To compare with experimental results, we employed a Metropolis-Hasting Monte Carlo algorithm\cite{1953Metropolis} based on approximating each island as a point dipole Ising spin governed by the Hamiltonian,
\begin{linenomath*}
\begin{equation*}
\pazocal{H}_{\text{dip}} = D \sum_{i \neq j} s_i s_j \bigg( \frac{\vec{\sigma_i} \cdot \vec{\sigma_j}}{r_{ij}^3} - \frac{3 (\vec{\sigma_i} \cdot \vec{r}_{ij})(\vec{\sigma_j} \cdot \vec{r}_{ij})}{r_{ij}^5}\bigg),
\end{equation*}
\end{linenomath*}
where $D = \frac{\mu_0 (M_S V)^2}{4 \pi a^3}$ is the dipole constant, and $\vec{r}_{ij} = \vec{r}_{j} - \vec{r}_{i}$ is a vector connecting the position of spin $i$ to spin $j$. 
The full dipolar sum was implemented for each simulation.
The spin at site $i$ has a magnetic moment, $\vec{s}_i = s_i \vec{\sigma_i}$, with  $s_i = \pm1$ the polarity of the spin; and $\vec{\sigma_i}$ a unit vector parallel to the long axis of island $i$. 
The orientation of $\vec{\sigma_{i}}$ is dependent on the rotation angle, $\vartheta$, of the tiling pattern, and the sub-lattice to which spin $i$ belongs.
A single Monte Carlo step (MCS) was taken to be $N$ single spin flips (where $N$ is  the number of spins in the array).
For systems with periodic boundary conditions, the lattice size used was $50 \times 50$ vertices (comprising 5100 islands). 
The MC data in Figs. \ref{quench_results}, \ref{melting_results}c, \ref{melting_results}d, and \ref{KZM_results} were averaged over 10, 1000, 5, and 10 independent iterations, respectively.
Results were verified up to lattice sizes of $100 \times 100$ vertices and no appreciable difference was seen.
The energy barrier to island flipping was taken to be $10\,D$  with a standard deviation of $10\%$ to account for disorder within the sample.
These values are consistent with those used in other work\cite{2012Budrikis}. \\

\textbf{Defect density in the Kibble Zurek Mechanism.} 
For a continuous second order phase transition, the equilibrium correlation length, $\xi$,  and equilibrium correlation time, $\tau$, diverge as

\begin{align*} 
	\centering
	\xi &=   (T-T_C)^{-\nu}; \\ 
	\tau &=  (T - T_C)^{-z\nu},
\end{align*}
where $v$ is the exponent associated with the correlation length, and $z$ is the dynamic critical exponent. The Kibble Zurek mechanism describes the dynamics of a system as the critical temperature, $T_C$, is traversed in time. Writing $\Delta T \equiv T - T_C$, we assume that the temperature can be varied linearly so that
\begin{linenomath*}
\begin{equation*}
	\Delta T (t) = \pazocal{R} t,
\end{equation*}
\end{linenomath*}
at time, $t$, for some rate, $\pazocal{R}$. Here, we use Monte Carlo steps as a proxy for time, so that $\pazocal{R}$ has units of $T \cdot \mathrm{MCS}^{-1}$. Equating the time to the critical point with the relaxation time yields a timescale, 
\begin{linenomath*}
\begin{equation*}
	t^* = \pazocal{R}^{\frac{-z\nu}{1+z\nu}},
\end{equation*}
\end{linenomath*}
commonly called the freeze-out time. Under the Kibble Zurek mechanism, there exists a region close to the critical point in which the order parameter no longer evolves adiabatically. The average correlation length at this freeze-out time is 
\begin{linenomath*}
\begin{equation*}
	\xi (t = t^*) = \pazocal{R}^{\frac{-\nu}{1+z\nu}}.
\end{equation*}
\end{linenomath*}
Assuming critical exponents for the 2-D Ising universality class ($\nu = 1$, $z \approx 2.1665$\cite{1996Nightingale}), the expected value for the exponent  is $\sim0.315$. We are also interested in the scaling behaviour of $\langle \sum | \nabla \times \mathbf{V} | \rangle$. In this scheme\cite{note5}, the curl of vertex moments is based on the sum and difference of the components of neighbouring vertex moments. The vertex moments are themselves simply the sum of the island moments in each vertex. Taking the absolute value introduces products of at most two spin components---exactly the same as in the two-point correlator. We thus expect $\langle \sum | \nabla \times \mathbf{V} | \rangle$ to scale similarly to the correlation length except with the opposite sign to reflect its nature as a density: $\langle \sum | \nabla \times \mathbf{V} | \rangle \sim \xi^{-1}$.


\begin{thebibliography}{10}
\bibitem{note2}
	\bibinfo{note}{For the case of a T$_2$ vertex, this definition is unambiguous:
		the island moments within each sub-lattice of the vertex are parallel. For
		the case of a T$_3$ vertex, while moments in one sub-lattice align parallel,
		the remaining two moments are aligned antiparallel. Nonetheless, the T$_3$
		vertex carries a net moment.}
\bibitem{note6}
	\bibinfo{note}{This implicitly assumes that the systems are not glassy enough
		to jam in a local energy minima.}
\bibitem{note5}
	\bibinfo{note}{Choosing a coordinate system in which the vertex moments lie in
		the ($x,y$)-plane, only the $z$-component of $\nabla \times \mathbf{V}$ is
		non-zero. Writing the vertex moment at a lattice point ($i,j$) as
		$\mathbf{V}^{(i,j)} = (V_{x}^{(i,j)},V_{y}^{(i,j)})$, the curl becomes
		\begin{equation*} \partial_x V_y - \partial_y V_x \bigg|_{\,\,(i,j)} \approx
		\frac{1}{2a} \bigg( V_y^{(i+1,j)} - V_y^{(i-1,j)} - V_x^{(i,j+1)} +
		V_x^{(i,j-1)} \bigg),\end{equation*} in terms of the vertex moments at the
		neighbouring lattice points.}
\bibitem{note4}
	\bibinfo{note}{In fact, it can be shown analytically that the two tilings tend
		to slightly different limits in the paramagnetic regime: \begin{equation*}
		\lim_{T\to\infty} \sum | \nabla \times \mathbf{V} | = \begin{array}{ll}
		\frac{674+223 \sqrt{2}}{1024} &\approx \,\,0.966 \text{ vertex}^{-1} \,\,\,\,
		\text{ for }\vartheta = 0^{\circ} \\ \frac{15(181 + 248 \sqrt{2})}{8192}
		&\approx \,\,0.974 \text{ vertex}^{-1}\,\,\,\, \text{ for }\vartheta =
		45^{\circ}. \end{array} \end{equation*}}
\end{thebibliography}

\section*{Acknowledgements}

This work was supported by the Engineering and Physical Sciences Research Council (EPSRC grant nos. EP/L002922/1, EP/L00285X/1, EP/M024423/1, and EP/P001483/1). G.M.M. is supported by the Carnegie Trust for the Universities of Scotland. Y.L. was funded by the China Scholarship Council. R.M. is supported by the Leverhulme Trust. R.L.S. acknowledges the support of the Natural Sciences and Engineering Research Council of Canada (NSERC)---R.L.S. a \'et\'e financ\'e par le Conseil de recherches en sciences naturelles et en g\'enie du Canada (CRSNG). G.M.M. thanks Aurelio Hierro-Rodriguez for useful input during sample optimisation.

\section*{Contributions}

G.M.M. and Y.L. optimised the sample deposition. 
Y.L. performed atomic force microscopy on the samples. 
G.M.M. and G.W.P. performed the TEM experiments and processed the Fresnel data. 
G.M.M. with the assistance of R.M. performed Monte Carlo simulations. 
G.M.M., G.W.P., and R.M. wrote the manuscript. 
All other authors commented on the manuscript. 
S.McV and R.L.S. supervised the project. 

\section*{Additional Information}
The authors declare no competing interests.

\end{document}